\documentclass[preprint]{aastex}
\shorttitle{MULTIPLE SCATTERING POLARIZATION OF T-DWARFS}
\shortauthors{SENGUPTA AND MARLEY }
\received{2009 July 20}
\begin{document}

\title{MULTIPLE SCATTERING POLARIZATION OF SUBSTELLAR-MASS OBJECTS: T- 
DWARFS}

\author{Sujan Sengupta}
\affil{Indian Institute of Astrophysics, Koramangala 2nd Block,
Bangalore 560 034, India; sujan@iiap.res.in}

\and

\author{Mark S. Marley}
\affil{NASA Ames Research Center, MS-245-3, Moffett Field, CA 94035,  
U.S.A.;
Mark.S.Marley@NASA.gov}

\begin{abstract}

 While there have been multiple observational programs aimed at detecting
linear polarization of optical radiation emitted by  ultracool dwarfs,  
there has been comparatively less rigorous theoretical analysis of the  
problem. The general expectation has been that the atmospheres of  
those substellar-mass objects with condensate clouds would give rise
to linear polarization due to scattering. Because of rotation-induced
non-sphericity, there is expected to be incomplete cancellation of
disk-integrated net polarization and thus a finite polarization. For
cloudless objects, however, only molecular Rayleigh scattering will
contribute to any net polarization and this limit has not been well
studied.  Hence in this paper we present a detailed multiple
scattering analysis of the polarization expected from those T-dwarfs  
whose spectra  show absence of condensates.
For this, we develop and solve the full radiative transfer equations for
linearly polarized radiation. Only atomic and molecular Rayleigh  
scattering  are considered to be the source of polarization.
We compute the local polarization at different angular
directions in a plane-parallel atmospheres calculated for the range of
effective temperatures of T dwarfs and then average over the whole  
surface of the object. The effects of gravity and limb darkening as well
as rotation  induced non-sphericity are included. It is found that the
amount of polarization decreases with the increase in effective temperature.
It is also found that significant polarization at any local point in the
atmosphere arises only in the optical (B-band). However, the disk integrated
polarization--even in the B-band--is negligible.  Hence we conclude that,
unlike the case  for cloudy L dwarfs, polarization of cloudless T-dwarfs by 
atomic and molecular  scattering may not be detectable.  In the future we
will extend this work to cloudy L- and T-dwarf atmospheres.
\end{abstract}

\keywords{stars: low mass, brown dwarfs --- polarization, extinction  
--- scattering --- stars : atmospheres}

\section{INTRODUCTION}

  Brown dwarfs have masses sufficient to ignite deuterium burning but
insufficient to enter into the hydrogen burning main sequence.
Instead, they radiate away the gravitational potential energy produced
during their formation and contraction.  As an individual
brown dwarf cools it progressively passes through spectral types  
ranging from
late M through the L and T sequences.  Helpful reviews of brown dwarf
science may be found in  Chabrier \& Baraffe (2000),
Burrows et al. (2001), Kirkpatrick (2005), and Marley \& Leggett (2008).

While condensates such as iron, forsterite ($\rm Mg_2SiO_4$), and  
enstatite
($\rm MgSiO_3$) are favored to be formed in the atmosphere of
relatively warm L dwarfs \citep{lunine86, tsuji96, lodders99, burrows00,
allard01}, in the atmosphere of mid-type and later T dwarfs, grains  
condense well bellow the
photosphere and are not an important opacity source \citep{ackerman01,  
tsuji02,
chabrier00, burrows01, kirkpatrick05, marley08}.
Synthetic spectra of model atmospheres lacking condensates fit well with
the observed spectra  of field T dwarfs later than about type T4
(with effective temperature $T_{\rm eff}\sim 1200\,\rm K$;  
\cite{stephens09}) suggesting the absence
of condensates in the visible atmosphere of these objects. At the same  
time, inclusion of condensate cloud of various species explains the spectra of
L dwarfs (e.g., Cushing et al. 2008).

  \cite{sengupta01} first predicted that scattering of light by
the condensates in the photopshere of L dwarfs would give rise to  a  
significant amount of linear polarization. This polarization would of course  
cancel out when integrated over a spherical surface of the star. However,  
analysis of high resolution spectra of L dwarfs confirms that these objects
are rapid rotator with the projected rotational velocity ranging from as
high as 90 to 10 $\rm kms^{-1}$ \citep{basri00,mohanty03, osorio06, reiner,
jones}.  A rapid rotation around its axis induces departure from sphericity
in the shape of the object.  As a consequence, the net polarization  
integrated over the disk does not completely cancel out.
Subsequently, linear polarization of several L dwarfs has been  
detected by various group of astronomers \citep{menard02, osorio05, goldman09,
tata09}. The observation were analyzed and explained by \cite{sengupta03,  
sengupta05} who assumed a single scattering atmosphere wherein the effect
of atoms and molecules was neglected. These authors suggested that scattering
by  atoms and molecules would give rise to negligible amount of polarization
and  hence polarization of T dwarfs in the optical should be inconsequential.   
However their single scattering approach left room for some uncertainty 
because they  considered a purely scattering medium while a proper approach
would incorporate both absorption and multiple scattering.

   In this paper we present the results of our detailed investigation  
on the polarization of T-dwarfs caused by multiple scattering of atoms and  
molecules in cloud-less atmospheres. In the next section we discuss the  
atmospheric models. In section~3, we present in detail the equations of
transfer of polarized radiation and in section~4 we present the numerical
methods of solving the transfer equation of polarized radiation in a
plane-parallel medium. The local polarization is integrated over the
rotationally  distorted surface.  We briefly describe the method adopted 
to calculate the rotational oblateness and the methods of integrating the
polarization over a rotationally distorted stellar disk in
section~5. We discuss the results in section~6 followed by our  
conclusion in the last section.

  \section{ATMOSPHERIC MODELS OF T-DWARFS}

The known field T dwarfs have effective temperatures in the range
$1400 > T_{\rm eff} > 550\,\rm K$.  Clouds influence the observed
spectra of the warmest dwarfs, but cloudless models generally
reproduce the spectra of most T dwarfs with $T_{\rm eff} < 1200\,\rm K
$, although this temperature is likely gravity dependent  
\citep{stephens09}.
  For the study reported here we considered only cloudless
models appropriate for spectral types later than about T3  
\citep{stephens09}.

For the polarization study we used four radiative-convective  
equilibrium models which give the
run of temperature as a function of pressure through the atmosphere  
from the collection employed by \cite{saumon08} in their study of brown
dwarf evolution.  The selected models include no cloud opacity, although
condensates are included in the chemical equilibrium calculation
\citep{freedman08}.  A more detailed summary of the model
atmospheres is presented in \cite{stephens09}.  The models shown
here all have $\log g = 5$ and solar metallicity.  The
effect of variation in these parameters will be explored in a
future study.  Here we choose models with $T_{\rm eff} = 1400$, 1200, 1000,
and 800 K.  As noted above, the first is somewhat warm for a typical
cloudless field T dwarf but we include it as a comparison point for  
future studies of cloudy L dwarfs with this $T_{\rm eff}$.

The four model temperature-pressure profiles described above were then
used as inputs for the polarization radiative transfer model.  The
same atmospheric opacities employed in the calculation of the
radiative-convective equilibrium models \citep{freedman08} were used
in both  calculations.  We note that in the calculation of these  
temperature profiles we computed fluxes in 180 separate spectral bins.
 Within each bin we computed the radiative transfer 8 times, once for
each of eight k-coefficient gauss points \citep{marley02}.  The details of the
underlying opacity  calculation are presented in \cite{freedman08}.   
For the polarization study we followed this same procedure and computed
monochromatic intensities at centers of the same 180 flux bins using
the same opacities.  Final computed quantities are the gauss weighted
sum of the eight independent radiative transfer calculations at each
of the 180 wavelength points.  We note that while the resulting
spectrum has somewhat low spectral resolution, polarization changes
slowly with wavelength and this resolution  is adequate for our
purposes.

\section{EQUATIONS OF TRANSFER OF POLARIZED RADIATION}

\subsection{Representation of polarized radiation by Stokes parameters}

In order to formulate the equations of transfer in a gaseous medium the
most convenient representation of polarized radiation is by a set of  
four
parameters called Stokes parameters. \cite{chandra60} first
introduced the Stokes parameters in the equation of radiative transfer
with a slight modification of Stoke's representation.
In an elliptically polarized beam, the vibrations of the electric and
the magnetic vectors in the plane transverse to the direction of  
propagation are such that the ratio of the amplitudes and difference in
phases of the components in any two directions at right angles to each
other are  absolute constants.
A regular vibration of this character can be represented by
\begin{eqnarray}\label{1}
\xi_l=\xi^{(0)}_{l}\sin(\omega t-\epsilon_{l}),
\xi_r=\xi^{(0)}_{r}\sin(\omega t-\epsilon_{r})
\end{eqnarray}
where $\xi_{l}$ and $\xi_{r}$ are the components of the vibration along
two directions $l$ and $r$ at right angles to each other, $\omega$ the
circular frequency of the
vibration, and $\xi^{(0)}_{l}$, $\xi^{(0)}_{r}$, $\epsilon_{l}$ and
$\epsilon_{r}$ are constants.
If the principal axes of the ellipse described by $(\xi_{l},\xi_{r})$  
are
in directions making angles $\chi$ and $\chi+\frac{1}{2}\pi$
to the direction $l$,
the equations representing the vibration take the simplified forms :
\begin{eqnarray}\label{2}
\xi_{\chi}=\xi^{(0)}\cos\beta\sin\omega t,
\xi_{\chi+\frac{1}{2}\pi}=\xi^{(0)}\sin\beta\cos\omega t
  \end{eqnarray}
where $\beta$ denotes an angle whose tangent is the ratio of the axes
of the ellipse traced by the end point of the electric (or magnetic)
vector, and the numerical values of it lies between 0 and $\frac{1} 
{2}\pi$.
The sign of $\beta$ is positive or negative according as the  
polarization
is right-handed or left-handed.

In  equation (\ref{2}),  $\xi^{(0)}$ denotes a quantity proportional to
the mean amplitude of the electric vector and whose square is equal
to the intensity of the beam:

\begin{eqnarray}
I=[\xi^{(0)}]^{2}=[\xi_{l}^{(0)}]^{2}+[\xi_{r}^{(0)}]^{2}=I_{l}+I_{r}.
\end{eqnarray}
Following the representation given in equation (\ref{2}) one obtains  
for the
vibrations in the $l$ and $r$ directions
\begin{eqnarray}
\xi_{l}=\xi^{(0)}(\cos\beta\cos\chi\sin\omega t-\sin\beta\sin\chi\cos 
\omega t)
\end{eqnarray}
and
\begin{eqnarray}
\xi_{r}=\xi^{(0)}(\cos\beta\sin\chi\sin\omega t-\sin\beta\cos\chi\cos 
\omega t).
\end{eqnarray}
The intensities $I_{l}$ and $I_{r}$ in the directions $l$ and $r$ can  
be written as
\begin{eqnarray}
I_{l}=[\xi_{l}^{(0)}]^2=I(\cos^{2}\beta\cos^{2}\chi+\sin^{2}\beta 
\sin^{2}\chi)
\end{eqnarray}
and
\begin{eqnarray}
I_{r}=[\xi_{r}^{(0)}]^2=I(\cos^{2}\beta\sin^{2}\chi+\sin^{2}\beta 
\cos^{2}\chi).
\end{eqnarray}

 From the previous equations it follows that if the regular vibrations
representing an elliptically polarized beam can be expressed in the form
given in equation (\ref{1}), we can write the relations
\begin{eqnarray}\label{8}
I=[\xi_{l}^{(0)}]^{2}+[\xi_{r}^{(0)}]^2=I_{l}+I_{r},
\end{eqnarray}
\begin{eqnarray}
Q=[\xi_{l}^{(0)}]^{2}-[\xi_{r}^{(0)}]^2= I\cos2\beta\cos2\chi=I_{l}- 
I_{r},
\end{eqnarray}
\begin{eqnarray}
U=2\xi_{l}^{(0)}\xi_{r}^{(0)}\cos(\epsilon_{l}-\epsilon_{r})=I 
\cos2\beta\sin2\chi=
(I_{l}-I_{r})\tan2\chi
\end{eqnarray}
and
\begin{eqnarray}\label{11}
V=2\xi_{l}^{(0)}\xi_{r}^{(0)}\sin(\epsilon_{l}-\epsilon_{r})=I 
\sin2\beta=
(I_{l}-I_{r})\tan2\beta\sec2\chi.
\end{eqnarray}
These are the Stokes parameters representing an elliptically polarized  
beam.
It follows from equation (\ref{8}) to equation (\ref{11}) that
$$ I^{2}=Q^{2}+U^{2}+V^{2}.$$
Further,
$$\tan2\chi=\frac{U}{Q}$$
and
$$\sin2\beta=\frac{V}{\sqrt{Q^{2}+U^{2}+V^{2}}}$$
which give the plane of polarization and the ellipticity respectively.

The nature of an arbitrary polarized light
is completely determined by the intensities
in two directions at right angles to each other (or equivalently, the  
total
intensity $I$ and $Q=I_{l}-I_{r})$ and the parameters $U$ and $V$. The
intensities $I$, $Q$, $U$ and $V$ are the general Stokes parameters
representing light.

\subsection{Rayleigh scattering}

In order to incorporate Rayleigh's scattering into the radiative  
transfer
equation, \cite{chandra60}  modified it which can be stated as the
vibrations representing the light scattered in a direction making an
angle $\Theta$ with the direction of incidence is
\begin{eqnarray}
\xi_{\parallel}^{(0)}=(\frac{3}{2}\sigma)^{1/2}\xi_{\parallel}^{(0)} 
\cos\Theta\sin(\omega t-\epsilon_{1})
\end{eqnarray}
and
\begin{eqnarray}
\xi_{\perp}^{(0)}=(\frac{3}{2}\sigma)^{1/2}\xi_{\perp}^{(0)} 
\sin(\omega t-\epsilon_{2}),
\end{eqnarray}
  where the phase $(\epsilon_{1},\epsilon_{2})$ and the amplitude $ 
(\xi^{(0)}_{\parallel},
\xi^{(0)}_{\perp})$, relations in the incident beam are maintained,  
unaltered, in the
scattered beam. Here $\parallel$ and $\perp$ refer to directions in  
the transverse
planes (of the incident and scattered light) parallel and perpendicular
respectively to the plane of scattering.
Accordingly, the parameters representing the scattered light are  
proportional to
\begin{eqnarray}
\frac{3}{2}\sigma[\overline{\xi^{(0)}_{\parallel}}]^{2}\cos^{2}\Theta= 
\frac{3}{2}
\sigma I_{\parallel}\cos^{2}\Theta,
\end{eqnarray}
\begin{eqnarray}
\frac{3}{2}\sigma[\overline{\xi^{(0)}_{\perp}}]^{2}=\frac{3}{2}\sigma  
I_{\perp},
\end{eqnarray}
\begin{eqnarray}
\frac{3}{2}\sigma[\overline{2\xi^{(0)}_{\parallel}\xi^{(0)}_{\perp} 
\cos(\epsilon_{1}-
\epsilon_{2})}]\cos\Theta=\frac{3}{2}\sigma U\cos\Theta
\end{eqnarray}
and
\begin{eqnarray}
\frac{3}{2}\sigma[\overline{2\xi^{(0)}_{\parallel}\xi^{(0)}_{\perp} 
\sin(\epsilon_{1}-
\epsilon_{2})}]\cos\Theta=\frac{3}{2}\sigma V\cos\Theta.
\end{eqnarray}

Therefore, denoting the incident light by the vector
$${\bf I}=(I_{\parallel},I_{\perp},U,V)$$
we can express the scattering intensity in the direction $\Theta$ by
\begin{eqnarray}
(\sigma\frac{d\omega '}{4\pi}){\bf RI}d\omega
\end{eqnarray}
where
$${\bf R}=\frac{3}{2}\left(\begin{array}{cccc} \cos^{2}\Theta & 0 & 0  
& 0\\
0 & 1 & 0 & 0\\0 & 0 & \cos\Theta & 0\\0 & 0 & 0 & \cos\Theta 
\end{array}\right)$$
${\bf R}$ is called the phase matrix for Rayleigh scattering.

\subsection{The explicit form of the phase matrix for Rayleigh  
scattering}

In the formulation of the equation of radiative transfer, the  
radiation field
at each point is characterized by the four intensities $I_{l}(\theta, 
\phi)$,
$I_{r}(\theta,\phi)$, $U(\theta,\phi)$ and $V(\theta,\phi)$ where $ 
\theta$
and $\phi$ are the polar angles referred to an appropriately chosen  
coordinate
system through the point under consideration and $l$ and $r$ refer to  
the
directions in the meridian plane and at right angles to it respectively.
Therefore one writes
$${\bf I}(\theta,\phi)=[I_{l}(\theta,\phi),I_{r}(\theta,\phi),U(\theta, 
\phi),
V(\theta,\phi)].$$
The explicit form of the phase function for Rayleigh scattering in  
terms of
$\theta$ and $\phi$ which is used in the transfer equation and which  
describes
the angular distribution of the radiation field is given by
(Chandrasekhar 1960):
\begin{eqnarray}
P(\mu,\phi;\mu',\phi') &=& Q[P^{(0)}(\mu,\mu')+(1-\mu^{2})^{1/2}(1- 
\mu'^{2})^{1/2}
P^{(1)}(\mu,\phi;\mu',\phi') \nonumber \\
& & + P^{(2)}(\mu,\phi;\mu',\phi')],
\end{eqnarray}
where
$$ P^{(0)}(\mu,\mu')=\frac{3}{4}\left(\begin{array}{cccc}2(1-\mu^{2}) 
(1-\mu'^{2})
+\mu^{2}\mu'^{2} & \mu^{2} & 0 & 0 \\\mu'^{2} & 1 & 0 & 0\\0 & 0 & 0 &  
0\\0 & 0
& 0 & \mu\mu'\end{array}\right) $$
$$ P^{(1)}(\mu,\phi;\mu',\phi')=\frac{3}{4}\left(\begin{array} 
{cccc}4\mu\mu'\cos(\phi'
-\phi) & 0 & 2\mu\sin(\phi'-\phi) & 0\\0 & 0 & 0 & 0\\-2\mu'\sin(\phi'- 
\phi) & 0
& \cos(\phi'-\phi) & 0 \\0 & 0 & 0 & \cos(\phi'-\phi)\end{array}\right) 
$$
$$ P^{(2)}(\mu,\phi;\mu',\phi')=\frac{3}{4}\left(\begin{array}{cccc} 
\mu^{2}\mu'^{2}
\cos2(\phi'-\phi) & -\mu^{2}\cos2(\phi'-\phi) & \mu^{2}\mu'\sin2(\phi'- 
\phi) & 0 \\
-\mu'^{2}\cos2(\phi'-\phi) & \cos2(\phi'-\phi) & -\mu'\sin2(\phi'- 
\phi) & 0 \\
-\mu\mu'^{2}\sin2(\phi'-\phi) & \mu\sin2(\phi'-\phi) & \mu 
\mu'\cos2(\phi'-\phi)
& 0 \\ 0 & 0 & 0 & 0\end{array}\right)$$
and
$$ Q=\left(\begin{array}{cccc}1 & 0 & 0 & 0\\0 & 1 & 0 & 0\\ 0 & 0 & 2  
& 0\\
0 & 0 & 0 & 2\end{array}\right)$$
Here $\mu=\cos\theta$, $\mu'=\cos\theta'$, $\theta$ and $\theta'$  
denote the
direction of the photon before and after scattering. Similarly for
$\phi$ and $\phi'$.

In case of the axial symmetry of the radiation
field, it clearly requires
that the plane of polarization be along the meridian plane (or, at  
right angle
to it). Consequently, $U=V=0$ and the two intensities $I_{l}$ and  
$I_{r}$
are sufficient to characterize the radiation field.

With this consideration the phase matrix used to describe the angular
distribution of photons that undergo Rayleigh scattering can be  
written as
\begin{eqnarray}
{\bf P}(\mu,\mu')=\frac{3}{4}\left(\begin{array}{cc}2(1-\mu^{2})(1- 
\mu'^{2})+
\mu^{2}\mu'^{2} & \mu^{2} \\ \mu'^{2} & 1 \end{array}\right)
\end{eqnarray}

\subsection{Equations for the transfer of polarized radiation in plane- 
parallel geometry}

In the present work, plane parallel atmosphere has been considered.
The transfer equation  governing the intensities $I_{l}$
and $I_{r}$ in rest frame with the medium stratified into plane parallel
can be written as
\begin{eqnarray}\label{eq1a}
\mu\frac{\partial}{\partial\tau}\left(\begin{array}{c}I_{l}(\mu)\\I_{r} 
(\mu)
\end{array}\right)=\left(\begin{array}{c}I_{l}(\mu)\\
I_{r}(\mu)\end{array}\right)-\frac{\omega_0}{2}\int^1_{-1}
{{\bf P}(\mu,\mu')\left(\begin{array}{c}I_{l}(\mu)\\I_r(\mu)\end{array} 
\right)
d\mu'}-(1-\omega_0){\bf b(z)}
\end{eqnarray}
where $\mu=\cos\theta (\mu\in[0,1])$ and $\theta$ is the angle between
the Stokes specific intensity vector and the axis of symmetry $z$;
${\bf b(z)}$ is the internal radiation source,  $\omega_0$ is the albedo
for single scattering and $\tau$ is the optical depth.

In case of spherical geometry the equation of transfer for polarized  
radiation can be written as
\begin{eqnarray}
&&\mu\frac{\partial}{\partial r}\left(\begin{array}{c}I_{l}(\mu)\\I_{r} 
(\mu)
\end{array}\right)+\frac{1-\mu^2}{r}\frac{\partial}{\partial\mu}
\left(\begin{array}{c}I_l(x,\mu,r)\\I_r(x,\mu,r)\end{array}\right)=
\nonumber \\ 
& & -\kappa\rho\left\{\left(\begin{array}{c}I_{l}(\mu)\\
I_{r}(\mu)\end{array}\right)-\frac{\omega_0}{2}\int^1_{-1}
{{\bf P}(\mu,\mu')\left(\begin{array}{c}I_{l}(\mu)\\I_r(\mu)\end{array} 
\right)d\mu'}-(1-\omega_0){\bf b(z)}\right\}
\end{eqnarray}
where $\kappa$ is the mass absorption co-efficient for radiation of  
frequency $\nu$ and $\rho$ is the density of the material such that
between the points $s$ and $s'$,

$$ \tau(s,s')=\int_{s'}^{s}{\kappa\rho} ds.$$

  For an object having radius as small as that of Jupiter with  
geometrically thin atmosphere, the second term in the left hand side of
the above equation which incorporates the curvature effect becomes
negligible. Therefore,  for solar type of stars and substellar-mass
objects, the plane-parallel stratification is sufficient.

   The degree of linear polarization p is given by
\begin{eqnarray}
p(\mu)=\frac{I_l(\mu)-I_r(\mu)}{I_l(\mu)+I_r(\mu)}
\end{eqnarray}

\section{NUMERICAL METHODS}

The work of \cite{redheffer62} and \cite{pre65} on the interaction
principle have been formalized by \cite{grant69}, Grant \& Hunt (1969a)  with
the introduction of the internal sources which is crucial for the
stellar atmospheres. Following this work, \cite{grant72} and
\cite{peraiah73} developed a method to obtain direct solution
of the transfer equations. This method is called the discrete space
theory of Radiative Transfer.
In the present work the method due to \cite{peraiah73} has been employed
to solve the radiative transfer equations in their vector form for  
linear polarization.

In this method, the entire medium is divided into N number of shells and
it is assumed that the specific intensities $I^{+}_{n}$ and $I_{n 
+1}^{-}$
are incident at the boundaries $n$ and $n+1$ respectively of the shell  
with
optical thickness $\tau$. The symbols with
signs + and - represent specific intensities of the rays travelling in
opposite directions.
If $\mu$ represents the cosine of the angle made by a ray with the
normal to the plane parallel layers
in the direction in which the geometrical depth decreases. That is,
$$I_{n}^{+}[I_{n}(\mu) : 0 < \mu \leq 1]$$
and
$$I_{n}^{-}[I_{n}(-\mu) : 0 < \mu \leq 1],$$
$I_{n}^{+}$ represents the specific intensity of the ray travelling in
the direction $\mu$ and $I^{-}_{n}$ represents the specific intensity
of the ray travelling in
the opposite direction. We select a finite set of values of
$\mu(\mu_{j} : 1 \leq j \leq m; 0 < \mu_1 < \mu_2 < \mu_3 \cdots \mu_m  
< 1)$
$$I^{+}_n=\left(\begin{array}{c}I_n(\mu_1) \\ I_n(\mu_2) \\ \vdots \\  
I_n(\mu_m)
\end{array}\right)$$
and
$$I^{-}_n=\left(\begin{array}{c}I_n(-\mu_1) \\ I_n(-\mu_2) \\ \vdots \ 
\ I_n(-\mu_m)
\end{array}\right)$$
are m-dimensional vectors on Euclidean space.

The incident intensity vectors are $I_{n}^{+}$ and $I_{n+1}^{-}.$  The  
emergent
intensity vectors are $I_{n}^{-}$ and $I_{n+1}^{+}.$
The emergent radiation field will have the contributions from the  
internal
sources, say, $\Sigma^{+}(n+1,n)$ and $\Sigma^{-}(n,n+1)$  
corresponding to
the output intensity vectors $I_{n+1}^{+}$ and $I^{-}_{n}$ respectively.

We assume certain linear operators which reflect and transmit the  
incident
radiation e.g., $t(n,n+1),$ $t(n+1,n),$ $r(n,n+1)$ and $r(n+1,n)$.
These operators are calculated based on the physics of the medium.
Then we can write the output intensities in terms of the transmitted and
reflected input intensities together with the internal sources as
\begin{eqnarray}\label{ieq1}
I_{n+1}^{+}=t(n+1,n)I_{n}^{+}+r(n,n+1)I_{n+1}^{-}+\Sigma^{+}(n+1,n)
\end{eqnarray}
\begin{eqnarray}\label{ieq2}
I_{n}^{-}=r(n+1,n)I_{n}^{+}+t(n,n+1)I_{n+1}^{-}+\Sigma^{-}(n,n+1).
\end{eqnarray}
The relationship given by equation (\ref{ieq1}) and equation  
(\ref{ieq2})
is called the Interaction Principle. Equation (\ref{ieq1}) and equation
(\ref{ieq2}) can be written concisely as
\begin{eqnarray}\label{ieq2a}
\left(\begin{array}{c}I^{+}_{n+1} \\ I_{n}^{-} \end{array}\right)=
{\bf S}(n,n+1)\left(\begin{array}{c} I_{n}^{+} \\ I^{-}_{n+1}  
\end{array}\right)
+\Sigma(n,n+1)
\end{eqnarray}
where
\begin{eqnarray}\label{ieq3}
{\bf S}(n,n+1)=\left(\begin{array}{cc}t(n+1,n) & r(n,n+1) \\ r(n+1,n)  
& t(n,n+1)
\end{array}\right)
\end{eqnarray}

If there is another shell with boundaries $(n+1,n+2)$ adjacent to (n,n 
+1),
interaction principle for this shell can be written as \citep{grant69}
\begin{eqnarray}\label{ieq3a}
\left(\begin{array}{c}I^{+}_{n+2} \\ I_{n+1}^{-} \end{array}\right)=
{\bf S}(n+1,n+2)\left(\begin{array}{c} I_{n+1}^{+} \\ I^{-}_{n+2}  
\end{array}\right)
+\Sigma(n+1,n+2)
\end{eqnarray}
where ${\bf S}(n+1,n+2)$ is similarly defined as in equation  
(\ref{ieq3}).
If we combine the two shells $(n,n+1)$ and $(n+1,n+2)$ then the  
interaction
principle for the combined shell is written as (for the thickness is  
arbitrarily
defined):
\begin{eqnarray}\label{ieq4}
\left(\begin{array}{c}I^{+}_{n+2} \\ I_{n}^{-} \end{array}\right)=
{\bf S}(n,n+2)\left(\begin{array}{c} I_{n}^{+} \\ I^{-}_{n+2}  
\end{array}\right)
+\Sigma(n,n+2).
\end{eqnarray}
${\bf S}(n,n+2)$ is called the star product of the two S-matrices
${\bf S}(n,n+1)$
and ${\bf S}(n+1,n+2)$; and ${\bf S}(n,n+2)$ can be written as
\begin{eqnarray}
{\bf S}(n,n+2)={\bf S}(n,n+1)\star{\bf S}(n+1,n+2).
\end{eqnarray}
Equation (\ref{ieq4}) is obtained by eliminating $I_{n+1}^{+}$ and  
$I_{n+1}^{-}$
from equation (\ref{ieq2a}) and equation (\ref{ieq3a}). We can write $r 
$ and $t$
operators for the composite cell as
\begin{eqnarray}\label{ieq5a}
t(n+2,n)=t(n+2,n+1)[I-r(n+2,n+1)r(n,n+1)]^{-1}t(n+1,n),
\end{eqnarray}
\begin{eqnarray}
t(n,n+2)=t(n,n+1)[I-r(n,n+1)r(n+2,n+1)]^{-1}t(n+1,n+2),
\end{eqnarray}
\begin{eqnarray}
r(n+2,n) &=& r(n+1,n)+t(n,n+1)[I-r(n+2,n+1)r(n,n+1)]^{-1}\times  
\nonumber \\
& & r(n+2,n+1),
\end{eqnarray}
\begin{eqnarray}\label{ieq5b}
r(n,n+2) &=& r(n+1,n+2)+t(n+2,n+1)[I-r(n,n+1)r(n+2,n+1)]^{-1}\times  
\nonumber \\
& & r(n,n+1)
\end{eqnarray}
and
\begin{eqnarray}
\Sigma(n,n+2)=\Lambda(n,n+1;n+2)\Sigma(n,n+1)+\Lambda'(n;n+1,n 
+2)\Sigma(n+1,n+2)
\end{eqnarray}
where $I$ is the identity matrix and
$$\Lambda(n,n+1;n+2)=\left(\begin{array}{cc}t(n+2,n+1)[I-r(n+2,n 
+1)r(n,n+1)]^{-1} & 0
\\ t(n,n+1)r(n+2,n+1)[I-r(n+2,n+1)r(n,n+1)]^{-1} & I \end{array}\right) 
$$
$$\Lambda'(n;n+1,n+2)=\left(\begin{array}{cc} I & t(n+2,n+1)r(n,n+1)[I- 
r(n,n+1)r(n+2,n+1)]^{-1}
\\ 0 & r(n,n+1)[I-r(n,n+1)r(n+2,n+1)]^{-1}\end{array}\right)$$
and
$$\Sigma(n,n+1)=\left(\begin{array}{c}\Sigma^{+}(n+1,n+2) \\
\Sigma^{-}(n,n+1)\end{array}\right).$$
Similarly, $\Sigma(n+1,n+2)$ is defined.

In order to obtain physical interpretation of the equations  
(\ref{ieq5a}) -
(\ref{ieq5b}) we expand the operator inverse in a power series. For  
example,
$$t(n+2,n)=\sum^{\infty}_{k=0}t_{k}(n+2,n),$$
$$t_{k}(n+2,n)=t(n+2,n+1)[r(n,n+1)r(n+2,n+1)]^{k}t(n+1,n). $$
This operator acts on intensities to the right and gives the  
contribution to
$I_{n+2}^{+}$ from $I_{n}^{+}.$
The term $t_{k}(n+2,n)$ may be recognized as diffuse transmission from  
n to n+1,
diffuse reflection from the layer $(n,n+1),$ k times in succession and  
finally diffuse transmission
through $(n+1,n+2)$. Thus $t(n+2,n)$ is the sum of contributions  
involving
scattering of all orders $k=0,1,2,\ldots,\infty.$  A similar
interpretation can be given for other operators.

If we write ${\bf S}(\alpha)$ ($\alpha$ to designate the cell) then
\begin{eqnarray}
  {\bf S}(\alpha\star\beta)={\bf S}(\alpha)\star{\bf S}(\beta)
\end{eqnarray}
where $\alpha\star\beta$ denotes the region obtained by putting the  
two cells $\alpha$
and $\beta$ together. If the cells are homogeneous and plane parallel  
then
\begin{eqnarray}
\alpha\star\beta = \beta\star\alpha.
\end{eqnarray}
In general, star multiplication is non-commutative. However, star
multiplication is associative.
If we have to add several layers $\alpha,\beta,\gamma,\ldots$ then,
\begin{eqnarray}
[(\alpha\star(\beta\star\gamma)\star\ldots)]={\bf S}[(\alpha\star\beta) 
\star\gamma\star\ldots].
\end{eqnarray}
If the medium is optically very thick then we can use what is known as
`doubling method'. For example,
\begin{eqnarray}
{\bf S}(2^{P}d)={\bf S}(2^{P-1}d)\star{\bf S}(2^{P-1}d),  
(P=1,2,3,\ldots)
\end{eqnarray}
which means that we can generate the S-matrix for a layer of thickness  
$2^{P}d$
in P cycles starting with ${\bf S}(d)$ rather than in $2^{P}$ cycles  
of adding
the ${\bf S}(d)s$ one by one. For example if $P=10,$ then only a  
fraction $10/2^{10}\simeq 10^{-2}$
of the computational work is needed to add $2^{10}$ layers of  
thickness d.

We expect the reflection and the transmission operators to be non- 
negative
on the physical grounds that intensities are always non-negative. This
condition will be satisfied only when the optical thickness of the shell
is less than certain value called the `critical size' or $\tau_{crit}$.
If the optical thickness $\tau$ of the shell in question is larger than
$\tau_{crit}$ then we can divide
the shell into several sub-shells whose thickness $\tau$ is less than
$\tau_{crit}$ and then use star algorithm to calculate combined response
from the sub-shells whose total thickness is T. If, for example, we need
the radiation field at internal points in the atmosphere, we shall have
to divide the entire medium into as many shells as we need and calculate
the radiation field at the N points in the medium. One can write down
the interaction principle for each shell and solve the whole system of
equations.

The solution $I^{+}_{n+1}$ and $I_{N}^{-}$ (for
any shell between shell 1 (at the outermost region) and shell N (at the
innermost region) are obtained from the relations
\begin{eqnarray}\label{eqi1}
I^{+}_{n+1}=r(1,n+1)I^{-}_{n+1}+V^{+}_{n+1/2}
\end{eqnarray}
and
\begin{eqnarray}\label{eqi1a}
I^{-}_{n}=t(n,n+1)I^{-}_{n+1}+V^{-}_{n+1/2}
\end{eqnarray}
with the boundary conditions $I^{-}_{N+1}=I^{-}(a).$
The quantities $r(1,n+1),$ $V^{+}_{n+1}$ and $V^{-}_{n+1}$ are  
calculated
by employing the initial conditions $r(1,1)=0$ and $V^{+}_{1/2}=I^{+} 
(b).$
The computation is done by the following recursive relations :
\begin{eqnarray}
& r(1,n+1)& =r(n,n+1)+t(n+1,n)r(1,n)[I-r(n+1,n)r(1,n)]^{-1}t(n,n+1),  
\nonumber \\
& & V^{+}_{n+1/2}=t(n+1,n)V^{+}_{n-1/2}+\Sigma^{+}(n+1,n)+R_{n 
+1/2}\Sigma^{-}(n,n+1),  \nonumber \\
& & V^{-}_{n+1/2}=r(n+1,n)V^{+}_{n-1/2}+T_{n+1/2}\Sigma^{-}(n,n+1),
\end{eqnarray}
where
\begin{eqnarray}\label{eqi1b}
t(n+1,n)=t(n+1,n)[I-r(1,n)r(n+1,n)]^{-1}, \nonumber \\
r(n+1,n)=r(n+1,n)[I-r(1,n)r(n+1,n)]^{-1}, \nonumber \\
R_{n+1/2}=t(n+1,n)r(1,n), \nonumber \\
T_{n+1/2}=[I-r(n+1,n)r(1,n)]^{-1}, \nonumber \\
t(n,n+1)=T_{n+1/2}t(n,n+1).
\end{eqnarray}

To calculate the radiation field at the internal points we proceed as  
follows :

\noindent
(1) Divide the medium into a number of shells (say N) with N+1  
boundaries as mentioned earlier. \\
(2) Start calculating the two pairs of reflection and transmission  
operators
$r(n+1,n)$, $r(n,n+1)$, $t(n+1,n)$ and $t(n,n+1)$ in each shell. If  
the optical thickness
of any shell is larger than $\tau_{crit}$ then apply star algorithm to  
use doubling procedure if
the medium is homogeneous. \\
(3) With the boundary condition that $r(1,1)=0$ and $V^{+}_{1/2}=I^{+} 
(b)$ and
the $r$ and $t$ operators mentioned in (2) compute recursively $r(1,n 
+1),$
$V^{+}_{1/2}$ and $t(n,n+1)$ given in equation (\ref{eqi1a}) to equation
(\ref{eqi1b}) from shell 1 to shell N i.e., from b to a.  \\
(4) Next sweep back from a to b calculating the radiation field given
in equation (\ref{eqi1}) with the boundary condition $I^{-}_{n+1}=I^{-} 
(a).$

   We use a eight point Gaussian quadrature formula for the numerical
integration over the angular points and hence $\mu$ and $\mu'$ are the
zeros of the Legendre polynomials of degree 8 in the interval
-1 to +1.  The phase matrix is normalized to 1.

\section{ROTATION-INDUCED OBLATENESS AND DISK INTEGRATED POLARIZATION}

   Once $I_l(\mu)$ and $I_r(\mu)$ are calculated in any local points,  
the
polarization integrated over the surface of the object can be calculated
which should be the observable quantity for point source of light.  
However,
the net polarization would cancel out to zero because of symmetry if the
the apparent disk of the object is perfectly spherical. Rapid
rotation of the object makes it non-spherical and hence non-zero  
polarization
arises when integrated over the surface.

The oblateness of a rotating object has been discussed by  
\cite{chandra33}
in the context of polytropic gas configuration under hydrostatic  
equilibrium.
For a slow rotator, the relationship for the oblateness $f$ of a stable
polytropic gas configuration under hydrostatic equilibrium is given by
\begin{eqnarray}\label{obl}
f=1-\frac{R_p}{R_e}=\frac{2}{3}C\frac{\Omega^2R^3_e}{GM}
\end{eqnarray}
where $M$ is the total mass, $R_e$ is the equatorial radius, $R_p$ is  
the polar
radius and $\Omega$ is the angular velocity
of the object. $C$ is a constant whose value depends on the polytropic  
index.
For the polytropic index $n=0$, the density is uniform and  $C=1.875$.
This configuration is known as the Maclaurin spheroid.
For a polytropic index of $n=1.0$, $C=1.1399$, which is appropriate for
Jupiter \citep{hubbard84}. For non-relativistic completely degenerate  
gas,
$n=1.5$ and $C=0.9669$.  Comparisons with detailed structure models
(D.~Saumon, private communication) show that brown dwarf interiors   
can be
adequately approximated by polytropes with $1 < n < 1.3$ with
the larger $n$ being appropriate for higher gravities.
For n=1 and $R_e=R_{\rm jup}$ the oblateness varies from 0.001 to  
0.086 when
the projected rotational velocity varies from V=10 to 90 ${\rm  
kms^{-1}}$
with the surface gravity $g=10^{5}$ ${\rm cms^{-2}}$ and it varies
from 0.00035 to 0.0287 with  $g=3\times10^{5}$ ${\rm cms^{-2}}$.  
Similarly,
for n=1.5 and  $R_e=R_{\rm jup}$, the oblateness varies from 0.0009 to
0.073 with  $g=10^{5}$ ${\rm cms^{-2}}$ and it varies from 0.0003 to  
0.024
with $g=3\times10^{5}$ ${\rm cms^{-2}}$ for the same variation in the
rotational velocity. It should be noted in this context that
at 1 bar pressure level, the oblateness $f$ of Jupiter,
Saturn and Earth are 0.065, 0.098 and 0.003 respectively.  
\citet{bar03} used
Darwin-Radau relationship and estimated the oblateness of
the exoplanet HD209458b to be about 0.00285 whereas the polytropic
approximation with n=1 yields a value of 0.00296.

Therefore, in the present work, we employ the
polytropic approximation of Chandrasekhar with the polytropic index  
$n=1$,
to find out the rotation-induced oblateness because this is  
appropriate for
low mass dwarfs and provides an upper limit to the oblateness and  
hence an
upper limit on the degree of polarization for a given rotational  
velocity.
As stated before, the surface gravity of all the models is fixed at
$g=10^{5}$ ${\rm cms^{-2}}$.

   We assume an axisymmetric shape of the object so that it has a  
rotational
invariance around some axis. The radius of any point on the surface is  
given
by
\begin{eqnarray}
R(\Omega,\mu)=\frac{R_e}{[1+(A^2-1)\mu^2]^{1/2}}
\end{eqnarray}
where A is the ratio of equatorial radius to the polar radius, i.e.,
$A=1/(1-f)$. Using the spherical harmonic expansion around $\theta$  
and $\phi$
\citep{jackson75}, we find the total flux integrated over the surface as
\begin{eqnarray}
F_I=\frac{1}{2}\sum_{l=0}^{\infty}(2l+1)P_l(\cos i)\int^1_{-1}\frac{P_l
(\mu)d\mu}{[1+(A^2-1)\mu^2]^{1/2}}\int^1_{-1}[I_l(\mu)+I_r(\mu)]\mu  
P_l(\mu)
d\mu
\end{eqnarray}
and the polarized flux as
\begin{eqnarray}
F_Q=2\pi\sum_{l=2}^{\infty}\alpha^2(l,2)P^2_l(\cos i) 
\int^1_{-1}\frac{P_l(
\mu)d\mu}{[1+(A^2-1)\mu^2]^{1/2}}\int^1_{-1}[I_l(\mu)-I_r(\mu)]\mu  
P^2_l(\mu)
d\mu
\end{eqnarray}

where
\begin{eqnarray}
\alpha(l,m)=\left[\frac{(2l+1)(l-m)!}{4\pi(l+m)!}\right]^{1/2},
\end{eqnarray}
$P_l(\mu)$ is the Legendre polynomial and $P_l^m(\mu)$ is the associated
Legendre polynomial. The disk integrated linear polarization is $p=F_Q/ 
F_I$.

The disk integrated polarization of a star is investigated by
\cite{harrington68} who assumed a Roche model wherein the entire
gravitational mass is considered to be at the center of the star. The
formalism of \cite{harrington68} although is valid for stars, is  
unlikely to be
applicable for brown dwarfs and giant gaseous planets because of their  
very
different mass distributions. Nevertheless, we use the formalism of
\cite{harrington68} and calculate the disk integrated polarization in  
order
to obtain an order of magnitude check in the results with spherical
harmonic expansion method. However, the
shape function $x(\Omega,\theta)$ in \cite{harrington68} is replaced by
\begin{eqnarray}
x(\Omega,\mu)=\frac{1}{[1+(1/A^2-1)\mu^2]^{1/2}}
\end{eqnarray}
so that the shape of the object takes the form of an ellipsoid, as  
before, and
the formalism becomes comparable with the spherical harmonic expansion  
method.
In both the methods we consider $R_e=R_{\rm jup}$ and the mass M  is  
derived
from the surface gravity which is fixed at $10^5$ cms$^{-2}$.

\section{RESULTS AND DISCUSSION}

   The effective temperature of T dwarfs varies from about 550K to  
1200K. Above 1200K, condensates of various species appears in the 
photosphere and transition from L dwarfs to T dwarfs takes place at
about 1300K to 1400K. On the other hand, T dwarf with effective
temperature less than 800K are likely too faint for image polarimetry. 
Therefore, in the present work we have considered an effective temperature
ranging  from 800K to 1200K. However, for a comparison with L dwarfs
of effective  temperature 1400K which have condensates in their photosphere
and hence should show detectable amount of polarization in the optical
because  of dust scattering, we have included the case for
$T_{\rm eff}=1400K$ without condensates. This
case provides a point of comparison to illustrate how
polarization differs with or without dust scattering.
The surface gravity of typical field brown dwarfs varies from  $10^5$  
to $3\times10^5$ cms$^{-2}$ (e.g., \cite{saumon08}). The departure
from rotation-induced sphericity decreases with the
increase in surface gravity \citep{sengupta05}. In order to maximize the
amount of polarization and hence to obtain an upper limit on the  
degree of polarization of T dwarfs, we have fixed  the surface gravity at
$g=10^5$ cms$^{-2}$.

   The numerical method used to calculate the local polarization  
provides unconditional stability for plane-parallel stratification of the  
atmosphere subject to the condition that each shell has less than the critical  
optical depth. This is achieved by dividing the shells with high 
optical depth  into sub-shells with optical depth less than the critical
optical depth.  Usually, $\tau<0.1$ serves the purpose. A convenient test
of the efficacy of the numerical method is to study the case of 
conservative scattering.  In a purely scattering medium, the physical system
must neither create nor destroy energy. For this purpose we apply
unpolarized incident radiation at the boundary of the inner radius of  
the shell and no radiation is incident at the outer shell.
For a conservatively scattering medium, the total flux that is  
introduced at the inner boundary is found to be equal within nine-th
decimal figure  to the sum of the flux that comes out of the outer
boundary and the  backscattered flux at the inner boundary. The polarization
is zero when the phase function $\bf P(\mu,\mu')=1$ i.e., when isotropic
scattering  is considered.

   Finally, the total emergent flux given by
\begin{eqnarray}
F=F_l+F_r=\int^1_{-1}{[I_l(\mu)+I_r(\mu)]\mu d\mu'}
\end{eqnarray}
matches exactly with the emergent flux obtained by \cite{marley00} and
\cite{stephens09} for scalar
radiative transfer case. In figure~1, we present the emergent flux for  
condensate-free atmospheres
with effective temperature ranging from 800K to 1400K. As it is
well known, the emergent flux in the optical falls rapidly with the  
decrease in wavelength owing to the presence of alkaline elements (Burrows,  
Marley \& Sharp 2000; Liebert et al. 2000). Therefore, T dwarfs
are extremely faint at shorter wavelengths.

   Figure~2(a-d) shows the polarization profiles at different  
directions for
different effective temperatures. The polarization is maximum when 
$\theta$, the angle between the direction of the radiation field and the  
symmetry axis is $90^o$.  However, the degree of polarization falls 
steeply with the decrease in $\theta$. At $\theta=90^0$, significant
polarization is  found even up to a wavelength of 1.5 ${\rm \mu m}$ for
the coolest T dwarf ($T_{\rm eff}=800K$).
But at a slightly smaller value of $\theta$, polarization becomes  
insignificant at wavelength longer than even 0.6 ${\rm \mu m}$.
The angular  dependency of the
polarization for different effective temperature is presented in  
figure~3.  As the angle between the direction of the radiation field
and the  symmetry axis decreases, the anisotropy in the radiation field
reduces drastically  because of the fact that at smaller angles, mostly
the unscattered photons  emerged
out and that the probability of scattering increases as $\theta$  
increases.  The probability of scattering is maximum at $\theta=90^o$.

   Figure~2(a-d) also shows that the degree of polarization decreases  
as the effective temperature of the object increases. In fact, this is true,  
not only at the outermost boundary but also at any depth inside the atmosphere.
In figure~4 we present the atmospheric depth dependence of the  
polarization at a particular wavelength ($\lambda=0.6\mu$m) and at a
direction $\mu=0.02$.  The degree of polarization for
$T_{\rm eff}=800K$ remains almost constant up to 1 bar of pressure  
level and then falls rapidly to zero. As the effective temperature 
increases, the degree of polarization falls rapidly to zero at a 
higher altitude or  at a lower pressure level. The physical explanation
of this feature becomes clear from
figure~5 wherein the variation of the single scattering albedo 
$\omega_0$ with respect to the atmospheric pressure P and temperature 
T is presented.  For
an object with $T_{\rm eff}=1200K$, the
scattering albedo is almost constant and its value is about one above a
pressure  level of $10^{-1}$ bar. Bellow this depth, it asymptotically
falls to zero. $\omega_0$ becomes zero at deeper region as one goes from
hotter to cooler objects. This means, the scattering albedo remains  
non zero up to much deeper region as the objects become cooler and hence  
contribution to the polarization originates from deeper region in cooler
objects.  As a consequence, the degree of polarization is higher in
cooler objects as  shown in figure~4.

   The polarization profile discussed above is calculated at a local  
point of the surface of the atmosphere stratified into plane-parallel
geometry.  This can be observable only if the object can be spatially
resolved.  However, a distant stellar object cannot be spatially resolved
and it appears to be a point source of light to an earth based observer.
Therefore only the disk averaged polarization is the observable quantity.
However, if the apparent disk of a stellar object is perfectly spherical,
because of symmetry, the net polarization averaged over the spherical 
disk would cancelled out.

   High resolution spectroscopic analysis of T dwarfs by \cite{osorio06,
del09} shows that just like L dwarfs, these objects are also fast  
rotators.  As a consequence, net non-zero polarization should arise
when integrated over the apparent disk because rotation induces
distortion in the  stellar disk.
In figure~6 and figure~7, we present the disk integrated polarization of
T dwarfs at an edge on view, i.e., at an inclination of $90^o$ with
rotational velocity V=90 and 60 kms$^{-1}$ respectively.
The degree of polarization is maximum when the inclination angle is
$90^o$, i.e., at an edge on view to an observer and it decreases with  
the decrease in the inclination angle as can be seen from figure~8.
\cite{harrington68} also reported the same feature. Figure~6 and  
Figure~7 shows that the degree of polarization of T dwarf with any
spectral type is non-zero only at wavelengths shorter than 0.6 ${\rm \mu m}$.
As the rotational velocity increases,
the oblateness of an object with the same mass, radius and polytropic
index increases. As a consequence with increasing rotational velocity  
there is less cancellation of polarization
over the surface,  yielding higher polarization. At the same time,
the effect of limb darkening and gravity darkening too increases and
consequently the degree of polarization increases further.
However, because of the fact that T dwarfs are extremely faint
at wavelengths shorter than 0.6 ${\rm \mu m}$, it would be extremely  
difficult to detect such a low polarization of T dwarfs, especially
if the  inclination angle is off the edge on view.

   The synthetic spectra calculated by using one dimensional plane- 
parallel atmospheric stratification match well with the observed spectra.  
However, a three dimensional approach is worth investigating in order to
achieve  better understanding of the effect of limb darkenning and the
polarization  integrated over the disk of a non-spherical object.

\section{CONCLUSIONS}

   In the present paper, we have discussed the results of our  
investigation on multiple scattering polarization of cloudless T dwarfs.
We predict that in the absence of condensates in the photosphere of 
T dwarfs later than about type T3 \citep{stephens09}, non-zero
polarization would arise only at wavelengths shorter than ${\rm 0.6 \mu m}$-
a spectral region where these dwarfs are exceptionally dark.
In order to calculate the maximum possible polarization of T dwarfs, we
have considered a polytropic equation of state with index $n=1$ which is
the minimum  permissible value of $n$ for representing the matter  
distribution in brown dwarfs. Consequently, the rotation-induced oblateness is  
maximized for a given rotational velocity.

We find that even with a rotational velocity as high as 90 kms$^{-1}$,  
the maximum amount of disk integrated polarization that arises at an  
inclination angle of $90^o$, is of the order $10^{-4}$, too small to
be detected  in the B-band.  The degree of polarization would be greatest
for the coolest objects and it would decrease with increasing  effective
temperature.  The results support the claim by \cite{sengupta03, sengupta05}
that the polarization detected in the optical region
(R- and I-band) of L dwarfs has negligible contribution from atomic and
molecular scattering. Therefore, if linear polarization in R- or I-band
of T dwarfs is detected in future, it will imply the presence of
condensates in the photosphere of T dwarfs.  Although a magnetic field
in principle might produce polarization,  T dwarf atmospheres are  
highly neutral (Gelino et al. 2002)
and thus the presence of magnetic field  would not give rise
to detectable amount of polarization in the optical \citep{menard02}.
Finally, polarization could play a
crucial role in measuring the loss of condensates from brown dwarf  
atmospheres as
we expect L- and early T-type dwarfs to show linear polarization in R-  
and I-band by dust scattering in the phototosphere
whereas mid-type and later T dwarfs would not show any polarization at  
wavelength greater than ${\rm 0.6 \mu m}$. The detailed investigation
of the polarization of L  dwarfs by
multiple dust scattering will be presented in a forthcoming paper.

\section{Acknowledgements}
S.S. is thankful to A. Peraiah for useful discussions. S.S. also  
acknowledges support by TIARA/ASIAA-National Tsing-Hua University, Taiwan
where a  part of this work was done. M.M. acknowledges support from the NASA
Planetary Atmospheres Program. Thanks are due to  the referee for constructive
comments.

\clearpage
\begin{figure}
\includegraphics[angle=0.0,scale=.70]{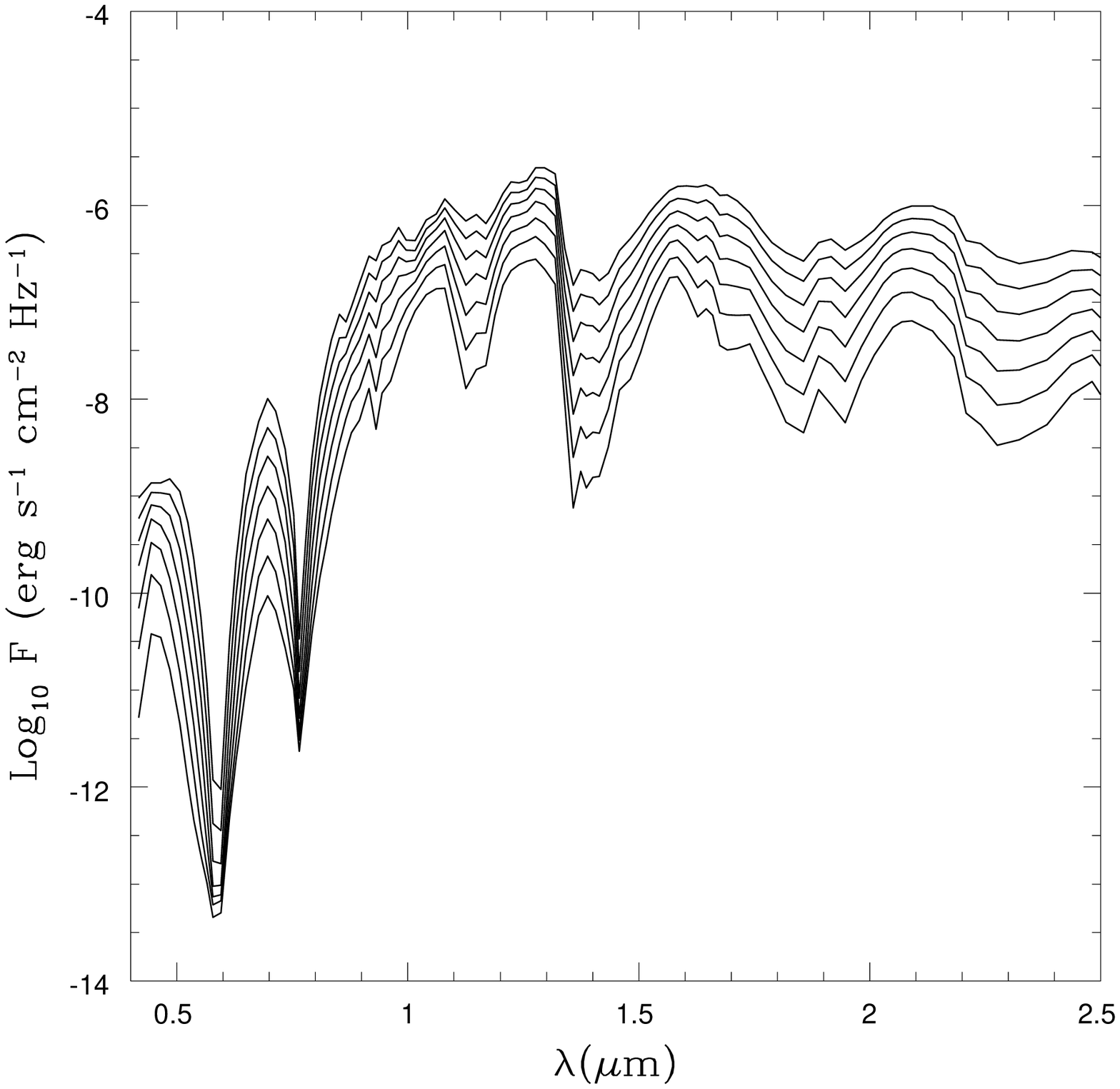}
\caption{The emergent flux as a function of wavelength for T dwarfs with
different effective temperature. From top to bottom, the solid lines  
represent the flux for $T_{eff}$=1400,1300,1200,1100,1000,900 and 800K  
respectively.
\label{fig1}}
\end{figure}

\begin{figure}\figurenum{2a}
\includegraphics[angle=0.0,scale=.70]{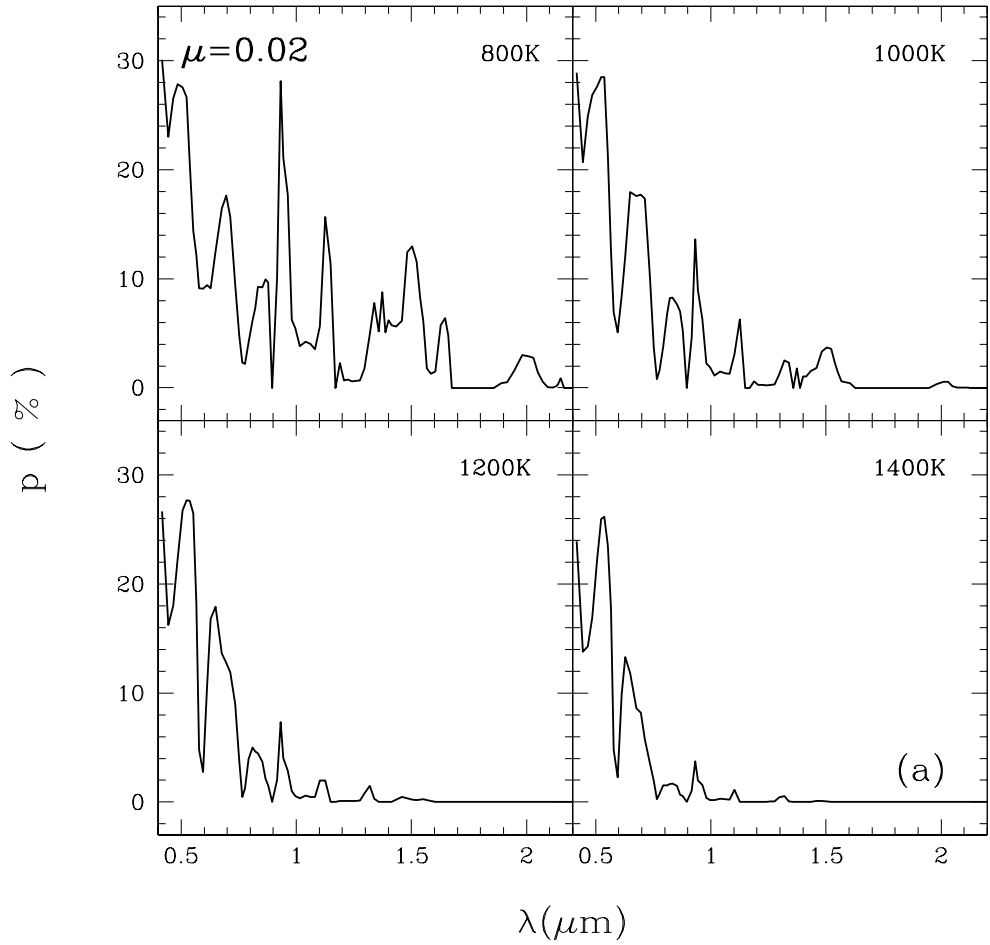}
\caption{Percentage degree of polarization along $\mu=0.02$ for T  
dwarfs with
effective temperature 800K, 1000K, 1200K and 1400K.
\label{fig2a}}
\end{figure}

\begin{figure}\figurenum{2b}
\includegraphics[angle=0.0,scale=.70]{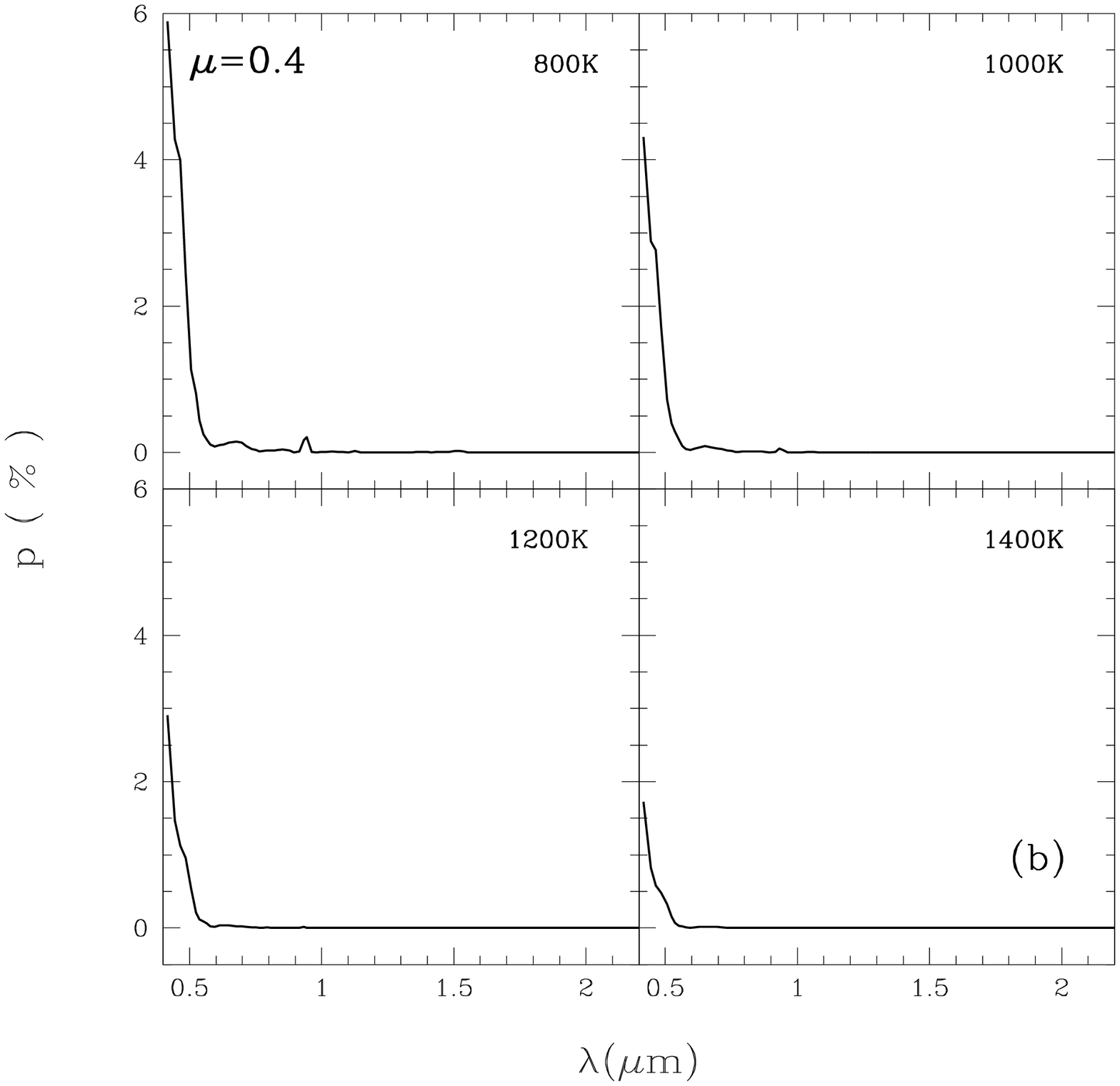}
\caption{Same as figure~2a but along $\mu=0.4$.
\label{fig2b}}
\end{figure}

\begin{figure}\figurenum{2c}
\includegraphics[angle=0.0,scale=.70]{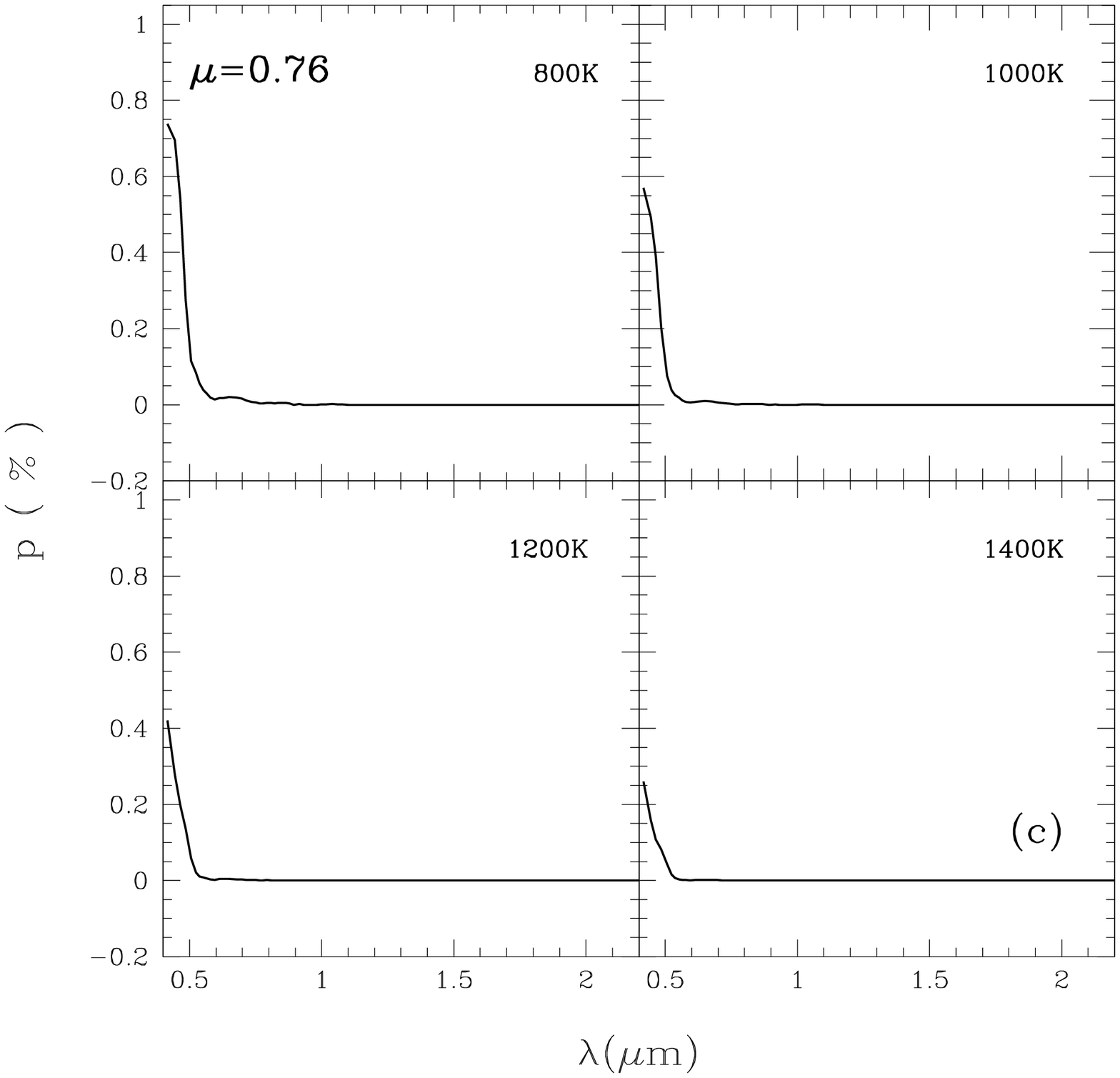}
\caption{Same as figure~2a but along $\mu=0.76$.
\label{fig2c}}
\end{figure}

\begin{figure}\figurenum{2d}
\includegraphics[angle=0.0,scale=.70]{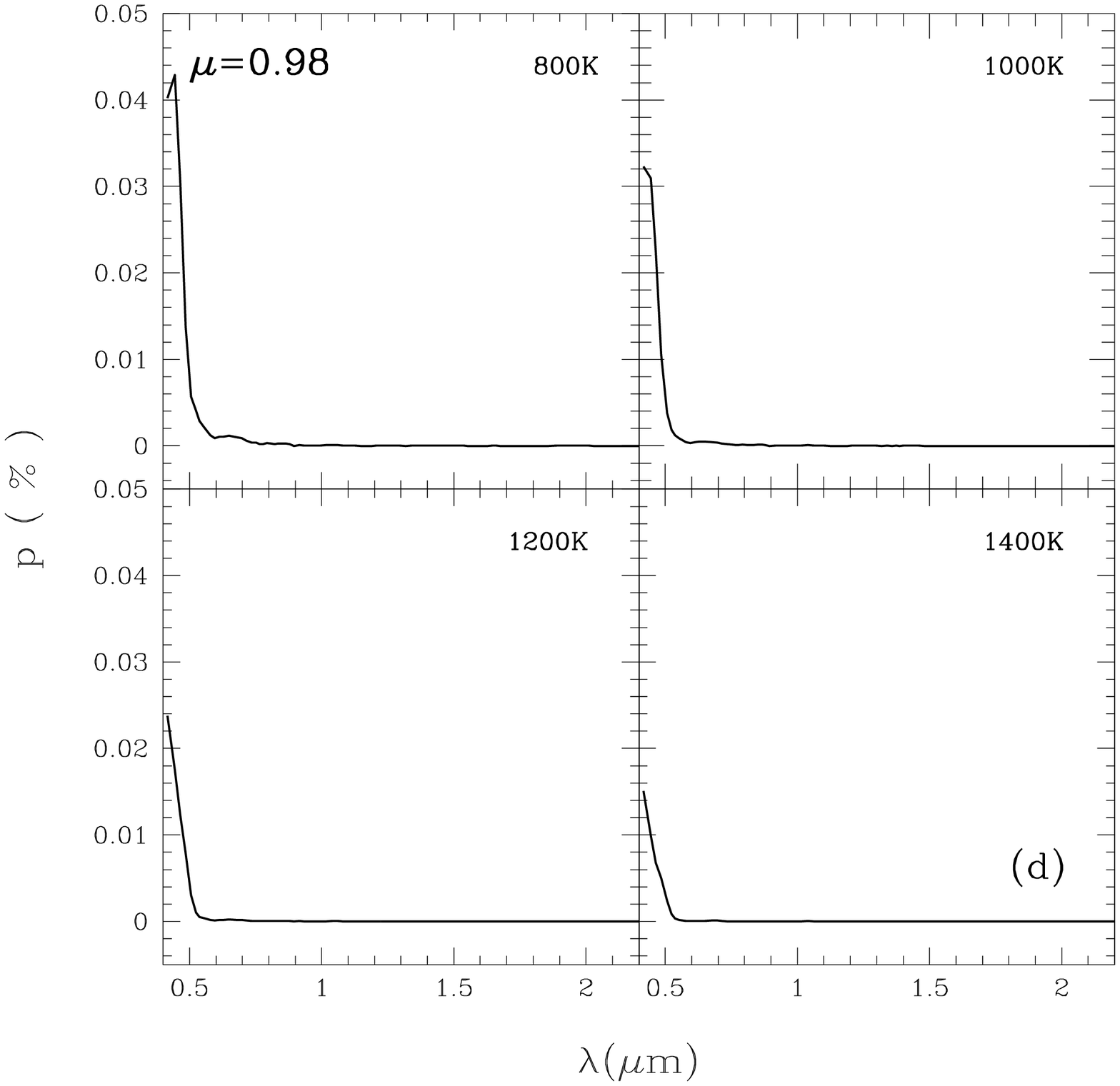}
\caption{Same as figure~2a but along $\mu=0.98$.
\label{fig2d}}
\end{figure}

\begin{figure}\figurenum{3}
\includegraphics[angle=0.0,scale=.70]{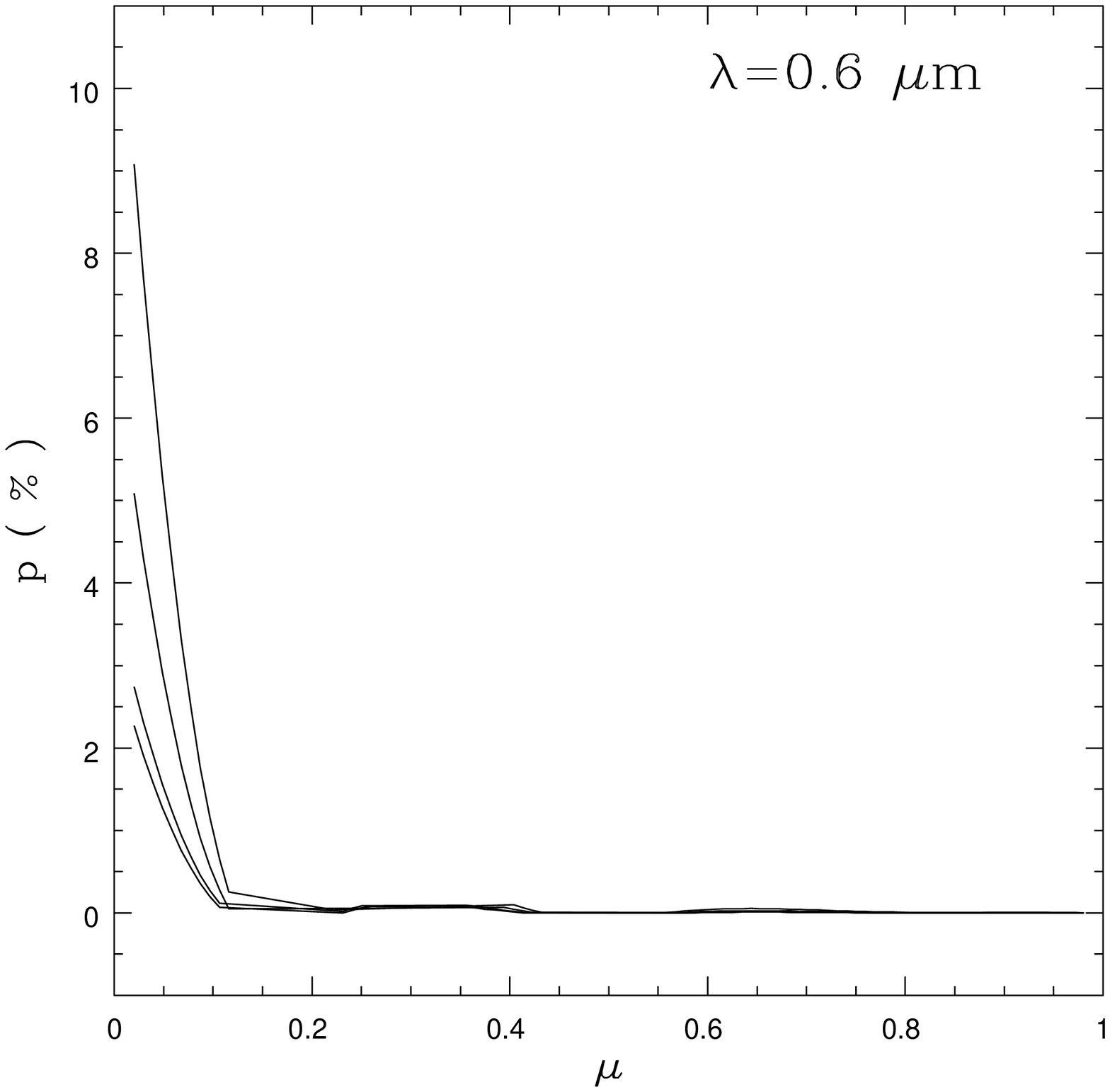}
\caption{Percentage degree of polarization at different $\mu=\cos\theta$ 
where $\theta$ is the angle between the direction of the radiation and the  
symmetry axis. The polarization is calculated at a wavelength $\lambda=0.6$
$\mu $m.
\label{fig3}}
\end{figure}

\begin{figure}\figurenum{4}
\includegraphics[angle=0.0,scale=.70]{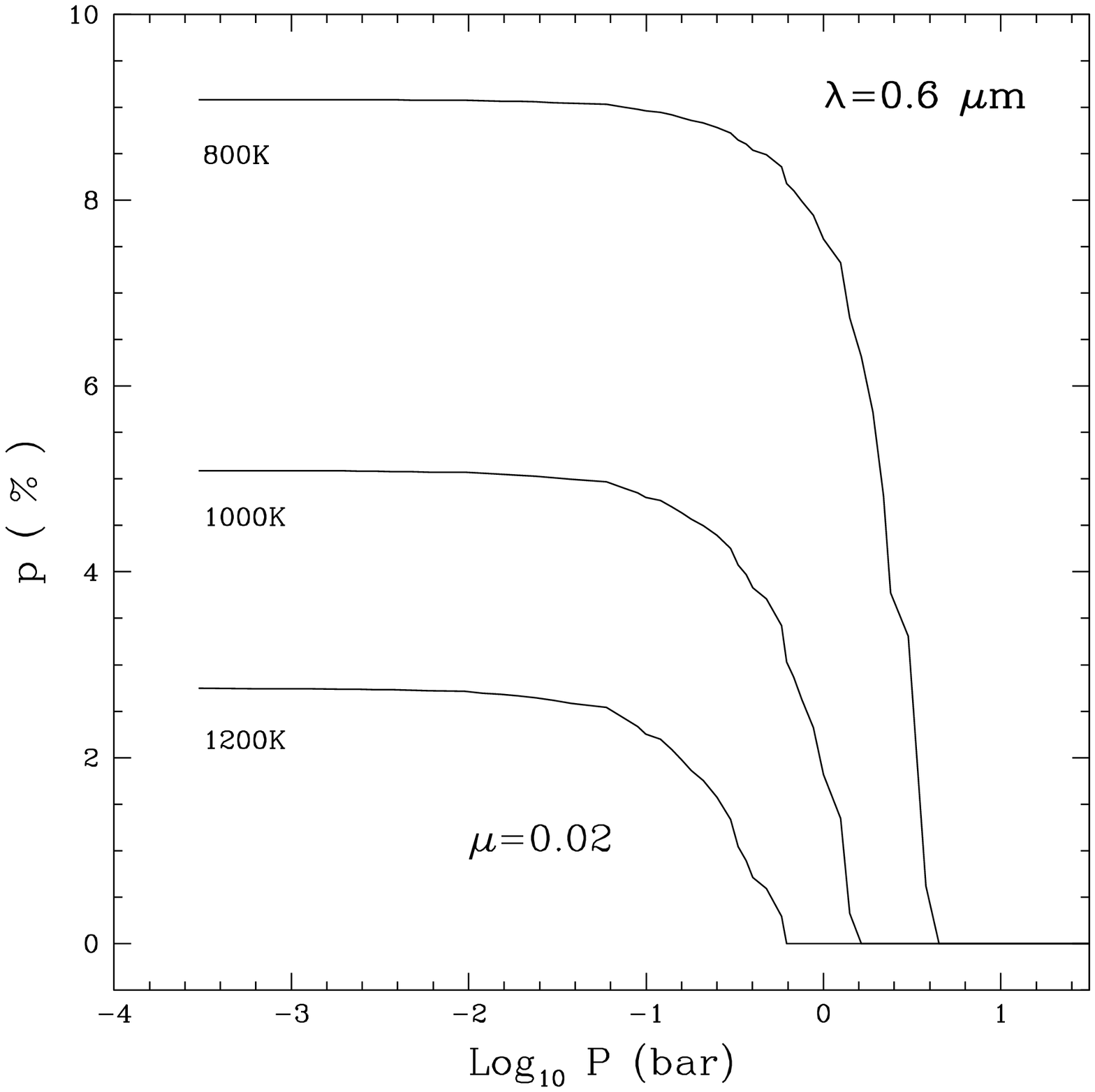}
\caption{Percentage degree of polarization along $\mu=0.02$ at different
atmospheric pressure level. The polarization is calculated at a  
wavelength $\lambda=0.6$ $\mu$m.
\label{fig4}}
\end{figure}

\begin{figure}\figurenum{5}
\includegraphics[angle=0.0,scale=.70]{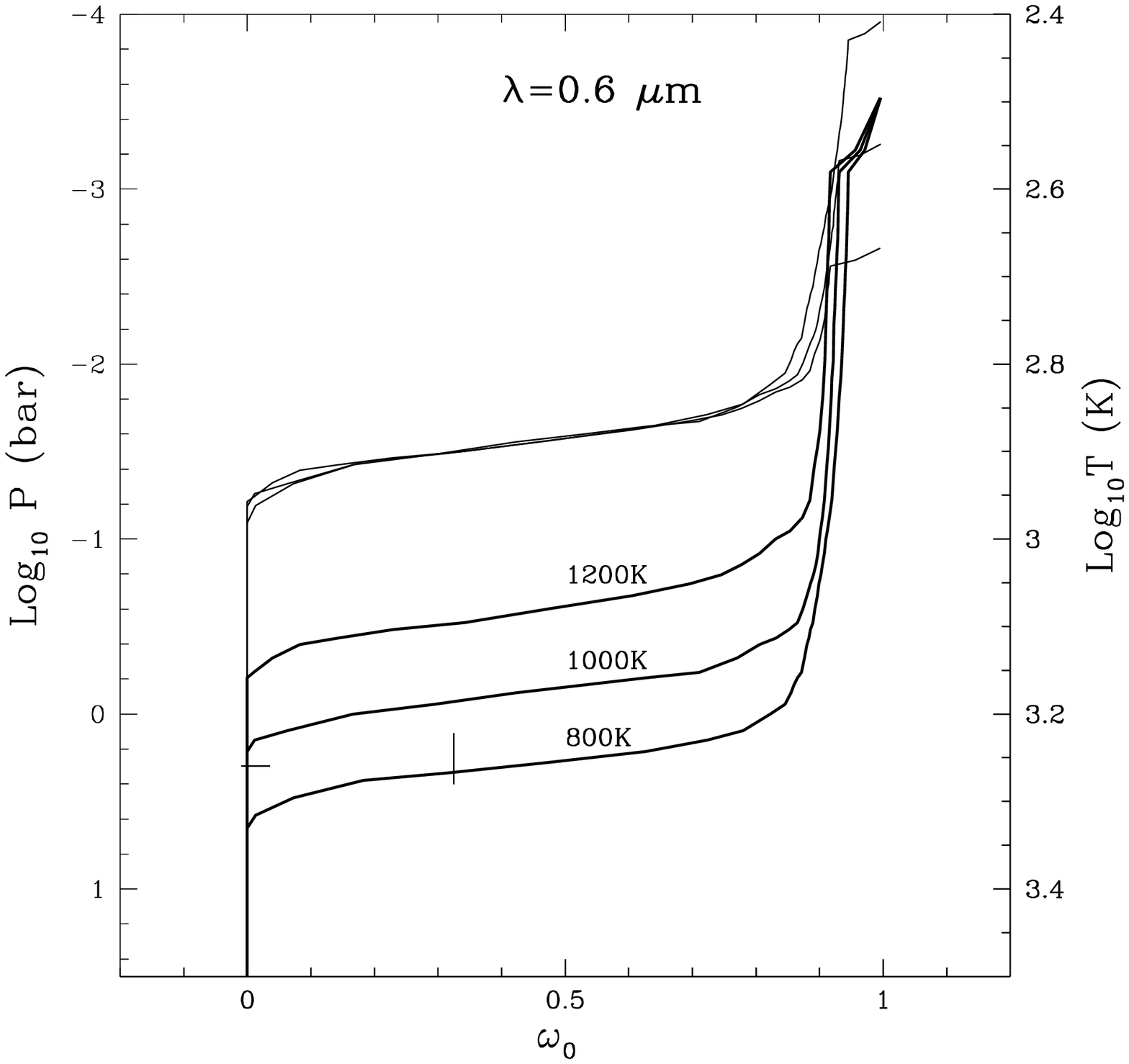}
\caption{Atmospheric depth dependence of single scattering albedo
$\omega_0$ at a wavelength $\lambda=0.6$ $\mu$m. The thick solid lines  
represent $\omega_0$ vs. atmospheric pressure P scaled in the left hand 
side axis.  The dashes over the lines indicate the location of the
photosphere.  The thin
solid lines represent the variation of $\omega_0$ with respect to the
atmospheric temperature T scaled in the right hand side axis.
\label{fig5}}
\end{figure}

\begin{figure}\figurenum{6}
\includegraphics[angle=0.0,scale=.70]{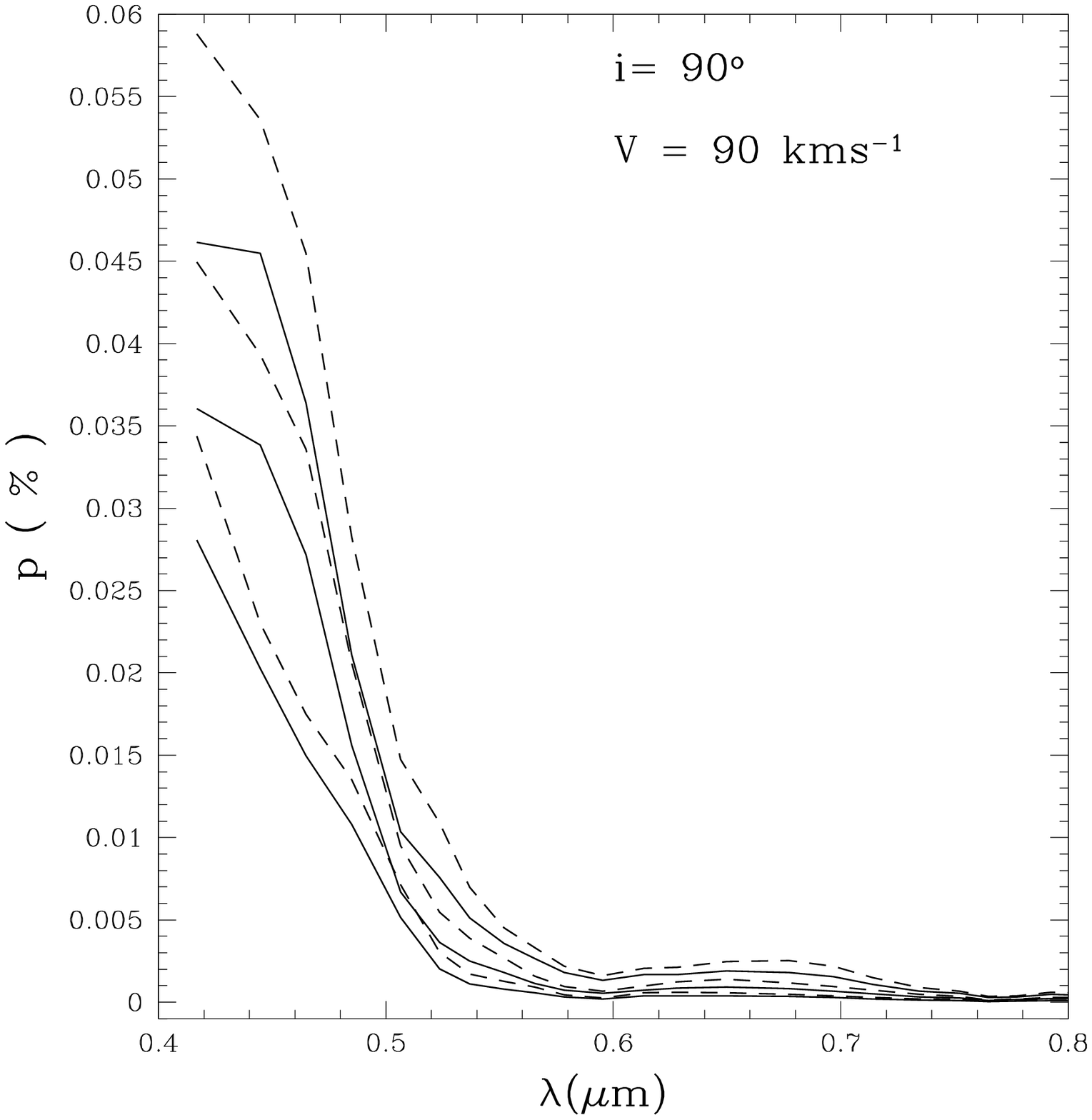}
\caption{The disk integrated polarization
at an inclination angle $i=90^o$ and for a rotational
velocity $V=90$ kms$^{-1}$. The solid lines represent the
percentage polarization calculated by using the spherical harmonic  
expansion method while the broken lines represent that calculated by using the
modified Harrington-Collins method. From top to
bottom the lines represent the disk integrated polarization of T  
dwarfs with effective temperature 800K, 1000K and 1200K respectively.
\label{fig6}}
\end{figure}

\begin{figure}\figurenum{7}
\includegraphics[angle=0.0,scale=.70]{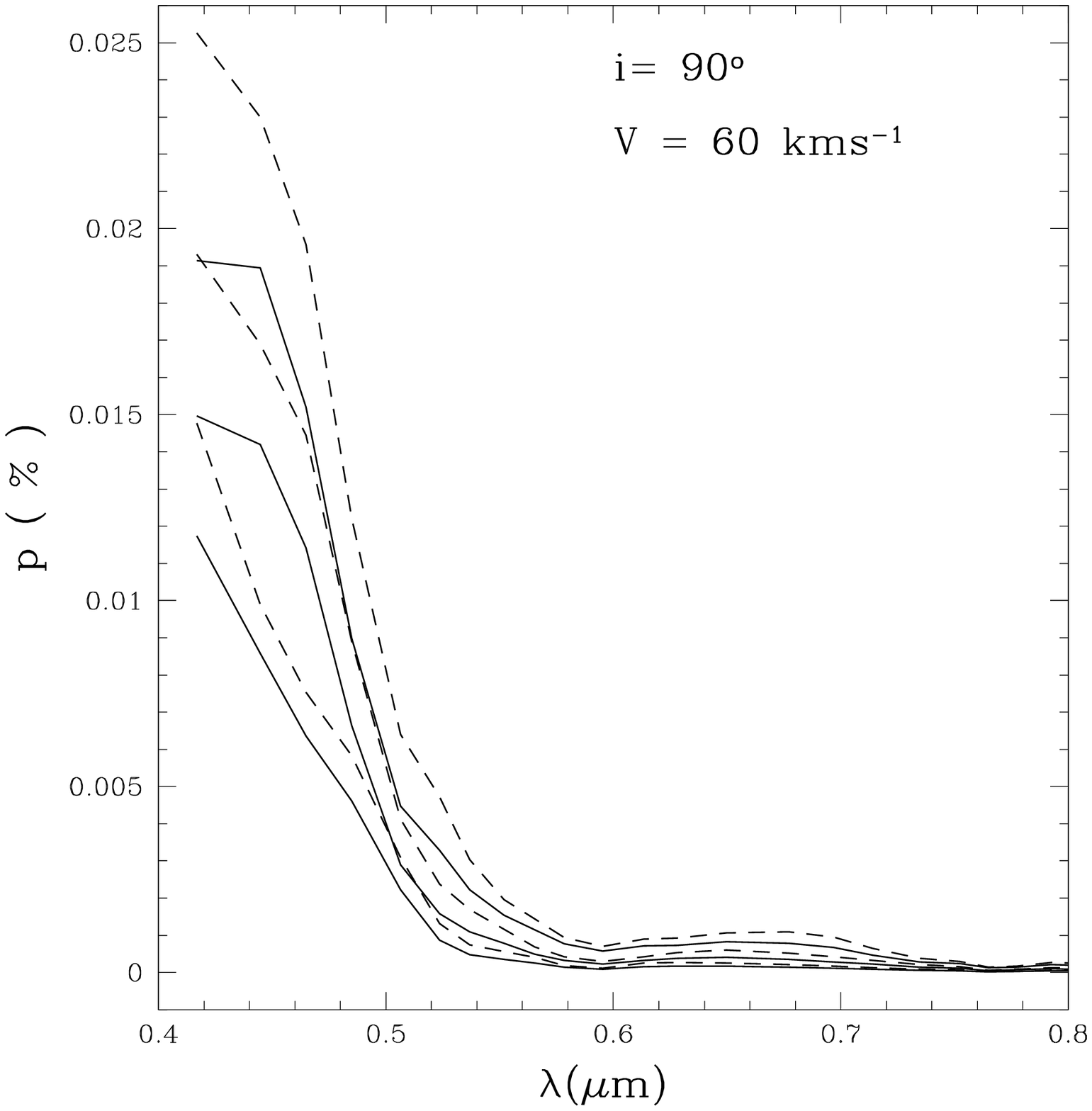}
\caption{Same as figure~6 but with $V=60$ kms$^{-1}$.\label{fig7}}
\end{figure}

\begin{figure}\figurenum{8}
\includegraphics[angle=0.0,scale=.70]{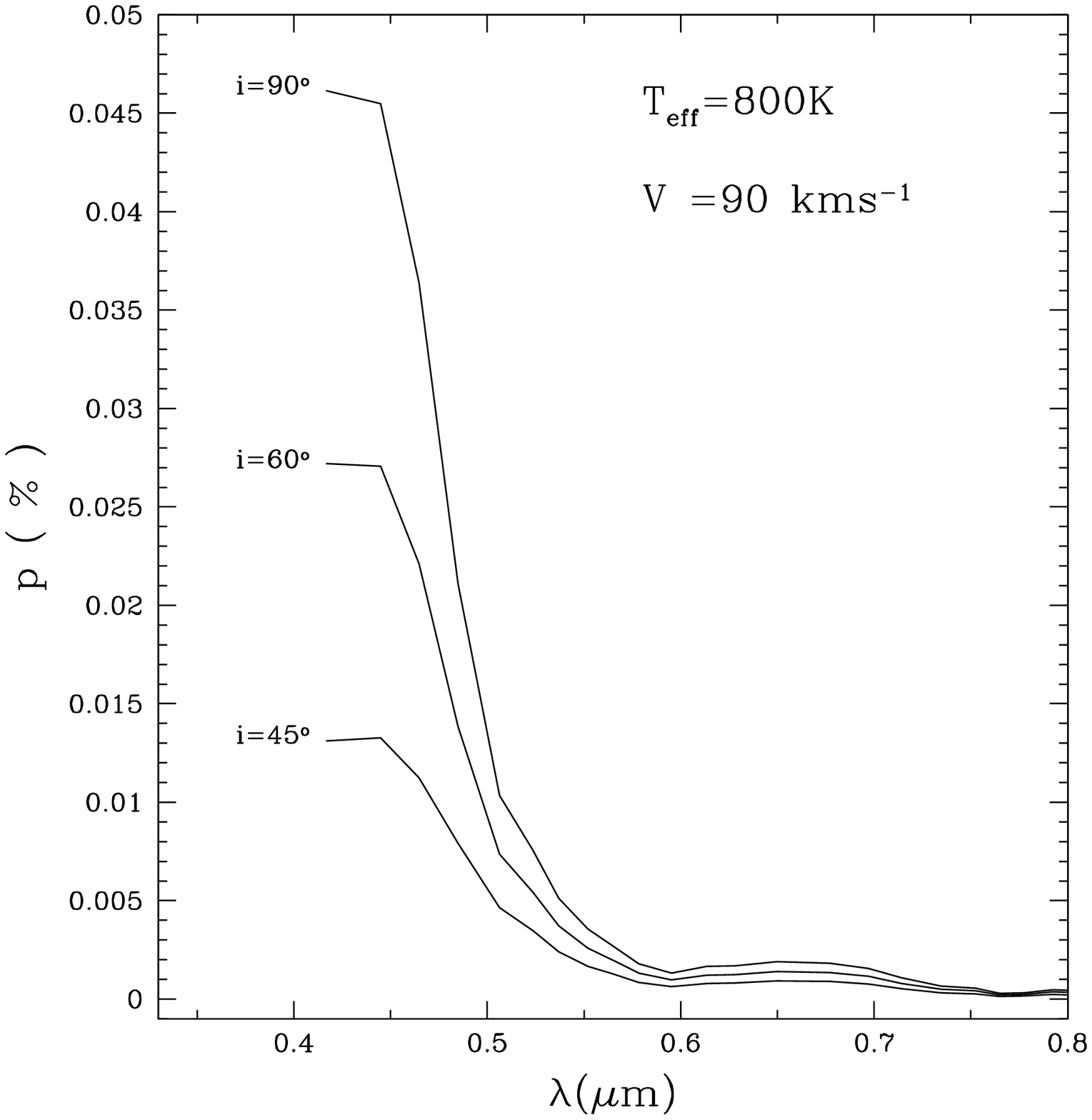}
\caption{Disk integrated polarization
at different inclination angles i. The polarization profiles shown  
here are calculated by using the spherical harmonic expansion method. 
\label{fig8}}
\end{figure}

\end{document}